# Observation of propagating edge spin waves modes

A. Lara[1], V. Metlushko[2], and F. G. Aliev[1*]

[1] *Dpto. Física de la Materia Condensada C-III, Universidad Autónoma de Madrid, 28049 Madrid Spain*

[2] *Dept. Electrical and Computer Engineering, University of Illinois, Chicago, 60607 IL, USA*

Broadband magnetization response of equilateral triangular 1000 nm Permalloy dots has been studied under an in-plane magnetic field, applied parallel (buckle state) and perpendicular (Y state) to the triangles base. Micromagnetic simulations identify edge spin waves (E-SWs) in the buckle state as SWs propagating along the two adjacent edges. These quasi one-dimensional spin waves emitted by the vertex magnetic charges gradually transform from propagating to standing due to interference and are weakly affected by dipolar interdot interaction and variation of the aspect ratio. Spin waves in the Y state have a two dimensional character. These findings open perspectives for implementation of the E-SWs in magnonic crystals and thin films.

(*) farkhad.aliev@uam.es

The possibility to excite and manipulate spin waves (SW) by magnetic fields or electric currents has opened perspectives for their implementation in communication and information processing and storage technologies [1-4]. Engineering of propagating and standing SWs in magnetic elements, as well as SW excitation and propagation in long strips or other two dimensional magnetic structures [3-5], has been a subject of recent intensive applied and fundamental research. Magnetic stripes with spin waves travelling inside are currently basic elements of magnonic waveguides [3,4].

Very recent studies predict however the possibility of excitation and propagation of a different kind of SW modes, so called edge spin waves (E-SWs) in individual two-dimensional magnetic structures [6,7] and extended magnonic crystals [8,9]. It has been suggested that the E-SWs are capable of providing new functionalities to spintronic and magnonic devices. These features include an easy modulation of SW spectra by mechanical structuring of the boundaries of the waveguide [6], unidirectional SW propagation, easily channelization, twisting, splitting, and manipulation, in a magnetic field perpendicular to the plane [8,9]. Spin waves have recently been reported to propagate along the long strip edges with inhomogeneous magnetization [10] when excited by perpendicular extended antenna. However, these spin waves were found to spread over the strip at high frequencies transforming into two-dimensional and losing therefore the edge character [10]. Observations of the E-SWs truly linked with the magnetically inhomogeneous edge states remain unclear.

Here we present experimental evidence of excitation and detection of quasi one-dimensional edge SWs emitted by the vertex magnetic charges of magnetic triangular dots. We observed that the edge SWs gradually transform from propagating to standing due to interference and are weakly affected by the variation of the dot aspect ratio. In order to effectively excite the spin waves locally one needs either to apply a strongly non homogenous local microwave field (which, for example, could be tailored by spin torque oscillator or small antenna [11-14] ) or by a quasi homogeneous microwave field interacting with non homogeneous local magnetization [15]. In our triangular dots the vertices are natural sources of strong local magnetic charges. The application of an in-plane magnetic field parallel to the triangles base, resulting in the buckle (B) state [16], peaks up homogeneously the exchange energy along the dot edges (other than the base) due to a competition between dipolar energy and exchange energies.

Previous interest in triangular magnetic dots [16-18] has recently been motivated by their possible implementation as magnetic logic elements [18] or versatile pinning centres for superconducting vortices [19]. These studies found that the influence of vertex topological magnetic charges diminishes with reduced lateral dimension [18] or with rounded corners [17]. As to the static magnetization distribution inside the dots, micromagnetic simulations and magnetic force microscopy [17,18,20] revealed three main topologically different magnetic states: vortex (V, with a vortex core in the triangle center), Y and B state. Magnetization dynamics has been studied by Brillouin light scattering and simulations, in sub-360nm equilateral triangles [20,21] with strongly suppressed vertex magnetic charges.

Our work investigates experimentally and by micromagnetic simulations the broadband response in 1000nm side length triangular dots which, being much larger than studied before [20,21], have well defined edge and vertex states. Micromagnetic simulations identify the modes excited in the B state as spin waves emitted by the vertex magnetic charges and propagating and interfering along the edge states. Remarkably, the observed E-SWs are qualitatively distinct from SWs observed previously in quasi one dimensional (1D) magnonic crystals [22], 2D spin wave diffraction [23, 24] or SW interference patterns from point contact spin torque emitters [12, 24-26].

The arrays of triangular Py dots were fabricated by a combination of e-beam lithography and lift-off techniques on a Si(100) substrate as explained elsewhere [27,28]. The dots are 30 nm thick, have a 1000 nm side length and a 200nm vertex-vertex and vertex-base separation between neighbours (Fig.1). The Py thickness of 30nm imposes the B or Y states and suppresses the low field V state in a broad range of applied in-plane DC magnetic fields ($H_{DC}$). Magnetic force microscopy and magnetization vs. field measurements (not shown) confirm that the vortex state appears only in the virgin samples. Once it is saturated, depending on the field orientation either B or Y states are created in a robust way. Static simulations show that low field V ground state is present in 1000nm side equilateral triangular Py dots with thickness exceeding 50nm.

Spin waves were excited with a coplanar waveguide (CPW, see sketch in Figure 1a) with a 30 μm wide central electrode and 20 μm gap to provide a nearly homogeneous in-plane excitation parallel to the triangles base. An about 30nm thick Si layer protects electrically Py dots from the CPW (sketch in Fig.1a). Two types of CPW allow aligning magnetic field parallel or perpendicular (B or Y states) to the microwave excitation, which is always directed along the triangles base. The spin wave spectra have been studied at room temperature by broadband vector network analyzer based VNA-FMR technique [27,28]. The measurements were made using a reflection configuration similar to the one described in [29], that only requires measuring the $S_{11}$ parameter, in which we fix $H_{DC}$ and sweep $S_{11}$ in frequency. To extract the dynamic magnetic response from the $S_{11}$ parameter we use a differential analysis in H, $U_d(H_i, f) = S_{11}(H_{i+1}, f) - S_{11}(H_i, f)$, that allows us to better observe field dependent eigenmodes, getting rid of the constant phase ripple in the VNA measurements and the constant background signal due to cables and connectors.

Besides the broadband measurements, static and dynamics micromagnetic simulations have been carried out using OOMMF [30]. To investigate numerically the magnetization dynamics we fix the magnetic state with $H_{DC}$ and then apply a short field pulse (1 ps full width half maximum, 5 Oe in amplitude) in the direction of the base (that we will call "x"). Doing this we reproduce the configuration of our experimental setup. By tracking the time evolution of the local magnetization in each simulation cell, we use Fourier Transform to get the excitation spectrum (in which peaks at certain frequencies indicate the presence of eigenmodes). The amplitude and phase of the Fourier Transform in each cell are put together to reconstruct the amplitude and phase profiles of oscillation of the whole dot for each single frequency. To reproduce the dispersion relation of the spin waves along a certain direction, 2D Fourier Transform is applied to the time evolution along the path containing the cells of interest, as will be shown below. We used the following parameters for Py:

exchange stiffness A = 1.4 ×10$^{-11}$ J/m, saturation magnetization $4\pi M_s$=10.43×10$^3$ G, Gilbert damping α = 0.01, and gyromagnetic ratio γ/2π = 2.96 MHz/Oe. The cell size used is 2.5nm x 2.5 nm in plane, and 30 nm in vertical direction.

Magnetic force microscopy data (Figure 1b) indicates the formation of the B state at zero magnetic field after saturating the dots in the "x" direction, which is in qualitative agreement with static simulations. A static simulation of a triangular dot in the same conditions is shown in Figure 1c, where colours represent the divergence of magnetization, usually compared to experimental MFM images [18]. Figures 1 d,e demonstrate that the direction of the applied magnetic field determines the formation of B or Y states. In each of these states edge domain walls (E-DWs) are formed in the close proximity to edges. The excess exchange energy in the B state has the unique topology of a nearly uniform 60 degree domain wall confined between the two vertex magnetic charges (Figures 1d).

Figure 2 presents the magnetic field dependence of magnetic dynamic permeability measured in the B (part a) and Y (part b) states, sweeping the static field from positive to negative values. For the B state well resolved spin wave modes show a continuous dependence on the magnetic field nearly down to small (100-200 Oe) field regions where a domain wall (in "y" direction) separating the dots in two halves is formed (Figure 1c). The observed eigenmodes that are roughly evenly spaced from each other correspond to E-SWs along E-DW. Given two consecutive eigenmodes, the one with higher frequency corresponds to a wave with an extra node of oscillation (compared to the eigenmode with lower frequency) between the two vertices that confine the E-SW (see below). The abrupt magnetization inversion of the triangular dots (not shown) can be noted in the nonzero negative field (-70 Oe) at which the resonances change from decreasing to increasing in frequency with the decreasing applied field. Apart from this, the modes are nearly symmetric at positive and negative fields. As for the spin waves in the Y state, some of lowest lying modes, however, show a qualitatively different dependence, with abrupt transformations of the modes with H$_{DC}$ (Figure 2b). Dynamic simulations qualitatively reproduce the main observation for the B state and remain practically unaffected by dipolar interactions (see Figure 2c). As we shall discuss further below, the interdot interaction is mostly relevant for the Y state where simulated eigenfrequencies for the single dot SW modes are qualitatively distinct from the experimental data (Fig. 2d). Some shift towards higher frequencies (of about 1GHz) observed in the simulations with respect to the measured eigenfrequencies could be due to the following factors: (i) the saturation magnetization of the Py triangles at room temperature could be smaller than the value used in the simulations done at zero temperature, (ii) the absence of defects in the simulations, and (iii) some difference in the value of damping.

The analysis of both magnitude and phase profiles of the main E-SW modes in the B state shows that in the field range (400<H<1500 Oe) they correspond to quasi one-dimensional (1D) spin waves emitted by the vertex magnetic charges, propagating along the two sides with E-DWs. Spin waves in the Y state have a mainly 2D character (Figure 2d, inset figure). Further below we shall mainly concentrate on understanding the nature of spin waves in the B state which reveal unexpected features. First, we note that these E-SWs gradually transform from propagating (near the vertices) to standing (close to the middle of the triangle side) due to interference of the spin waves with decaying amplitude, emitted by two coherent vertex emitters. Figure 3a presents a typical time evolution of the oscillations along a triangle edge in the B state, where the interference patterns reconstructed for different SW modes confirm a transformation from the propagating to standing E-SWs. Secondly, we have analyzed the dispersion of the E-SWs propagating along the left side of the triangle situated between vertices 1 and 3 the particular edge state (Figure 3b). We found that both the space-time and wavenumber-frequency representations point out some spatial E-SW asymmetry which could be linked with contribution of the secondary spin waves reflected from the vertical DW separating the dot in two halves, as well as the presence of this DW in vertex 3, while vertex 1 remains unaffected by this wall. Here the positive **k** vector branch corresponds to the SWs emitted by vertex 1 while the negative **k** branch describes E-SWs emitted by vertex 3 in the direction to vertex 1.

The predominantly 1D character of SWs along E-DW in the B state is also corroborated by their approximately parabolic dispersion relation (Figure 3b). Indeed, the exact solution for the spin waves in a 1D ferromagnetic chain predicts a parabolic $\omega \sim \mathbf{k}^2$ dispersion [31], valid except in the region of $|\mathbf{k}| \to 0$ where dipolar contribution dominate. The spin wave dispersion in one-dimensional bi-component magnonic crystal waveguides, as investigated by micromagnetic simulations [32] also shows a parabolic form, but is interrupted by the band gap due to imposed underlying periodicity. The dispersion curve of the spin waves excited along the non uniform E-DW in the Y state (Fig. 1e) is more complex.

The observed E-SW modes are qualitatively similar to recently reported string-like excitations along topologically pinned domain walls in the metastable double magnetic vortex state (the so called Winter's magnons ). The main difference is that in the present case the SW confinement is imposed by the sample boundary and the relative direction of the external magnetic field, while Winter's magnons propagate *inside* circular dots along topological domain walls pinned in between magnetic vortex cores and half edge antivortices [28].

We have also investigated numerically the possible influence of the variation in the aspect ratio of the triangular dots on the E-SWs along E-DWs in the B state. Dynamic simulations reveal qualitatively similar E-SWs excited in isosceles triangles with base and height equal to 1000nm (Fig. 3c). This demonstrates a sufficient stability of the E-SWs along the edge states to small variation of the lateral dot parameters.

Finally, in order to further corroborate the validity of spin wave identification (Fig. 2) we have analysed in more details the influence of dipolar interdot interactions on the observed E-SWs by comparing the distribution of exchange energy inside the dots and the stray fields obtained for single dots with those located inside a 4x4 dot array with the same spacing as those studied experimentally. Figure 4 summarizes the main observations for the B and Y states by representing the simulation results for a few selected fields. We have found that the topology of the E-DWs in the B state remains qualitatively unchanged when the interdot interaction is switched on (upper half in Fig 4). This results, added to the weak sensitivity of the main modes to dipolar interaction (Fig. 2c) and to variation of the aspect ratio (Fig. 3c) in the B state further manifest the robustness of the observed E-SWs. The edge states in the Y state (as well as excited spin wave modes, see Fig. 2), however, are much more strongly affected by dipolar interaction (see lower part in Fig.4). Seemingly, the main reason of the stronger influence of interactions in the Y state is the capability of a dot upper vertex to displace off the center the bottom E-DW unpinned magnetic charge of the dot above it, thanks to the stray field generated it generates. This possibility is evidently absent in the interacting B state dots which have only vertex (pinned and dipolar coupled) magnetic charges.

In conclusion, we have presented experimental evidence for spin waves propagating along edge states in triangular Py dots. These spin waves are shown to be robust with respect to variation of the dots shape from equilateral to isosceles dots with height equal to the base length. Although the large (few micron and above size) soft magnetic triangles could, in principle, be implemented for information transmission and processing using E-SWs, we hope that our findings will stimulate further experimental efforts in investigation the whole new class of topological spin wave modes propagating along the edges in different types of magnetic devices and creation of alternative ways for spin wave excitation, transmission and manipulation.

Authors acknowledge K. Guslienko for fruitful discussions. This work has been supported by the Spanish MINECO (MAT2012-32743, CONSOLIDER CSD2007-00010), Comunidad de Madrid (P2009/MAT-1726) and by the U.S. NSF, grant ECCS-0823813 grants. The authors acknowledge CCC-UAM for the computational capabilities (SVORTEX project). A.Lara thanks Universidad Autónoma de Madrid for FPI-UAM fellowship.

**Figures**

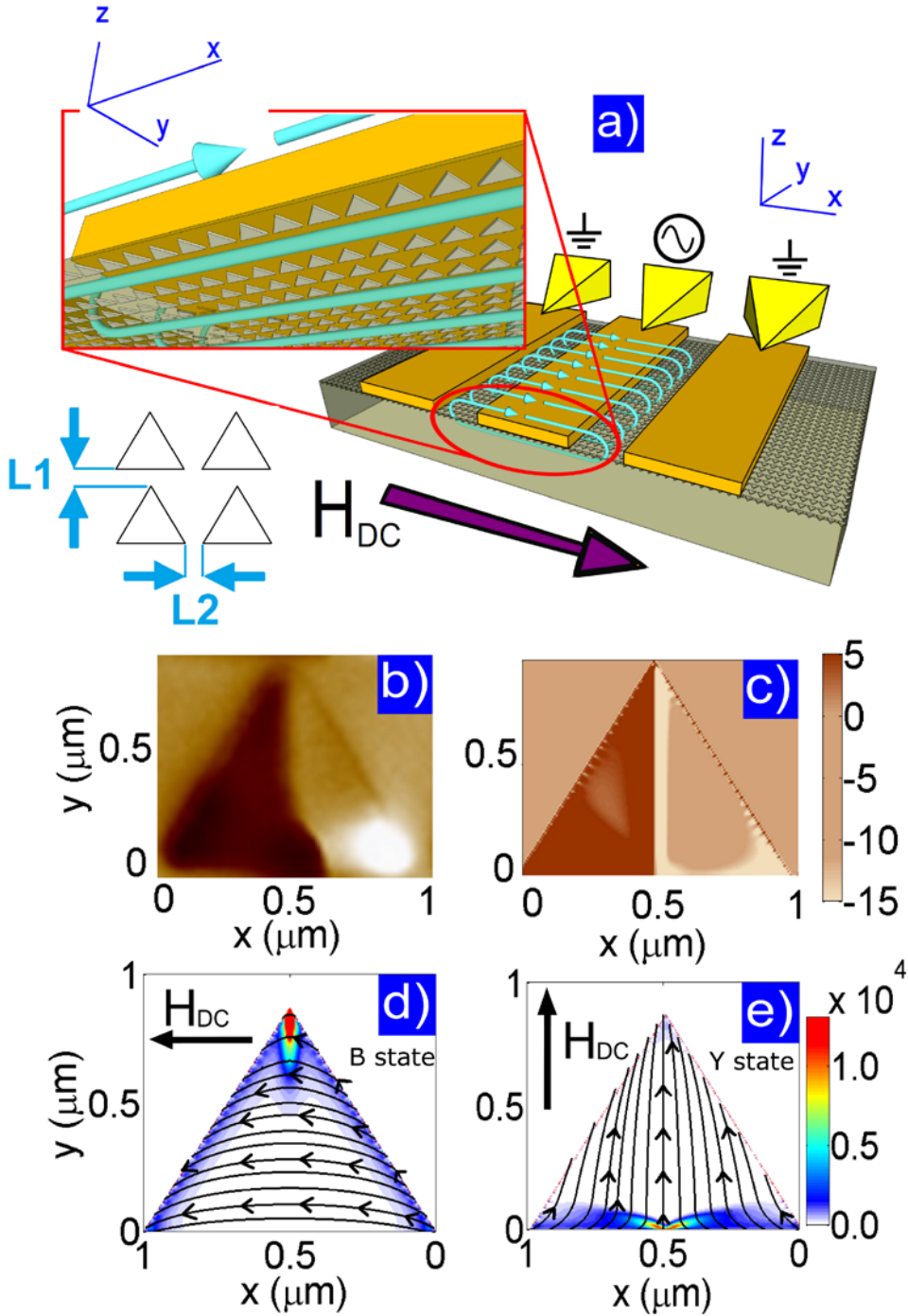

Figure 1. a) Sketch of the experimental system. VNA is connected to CPW with a high frequency ground-signal-ground probe. A high frequency current generates the rf field (blue arrows), incident on the triangles. A DC magnetic field ($H_{DC}$) can be applied in-plane, in the "x" (shown in sketch) or "y" direction. The interdot separations are L1=L2=200 nm. b) MFM image of a dot at zero field after saturating in "x" direction. B state is observed. c) Static simulation in the same field conditions. Colors represent the divergence of magnetization (a.u.). Static simulations of d) B and e) Y states. Arrows represent the direction of magnetization, and colors the density of exchange energy, in $J/m^3$.

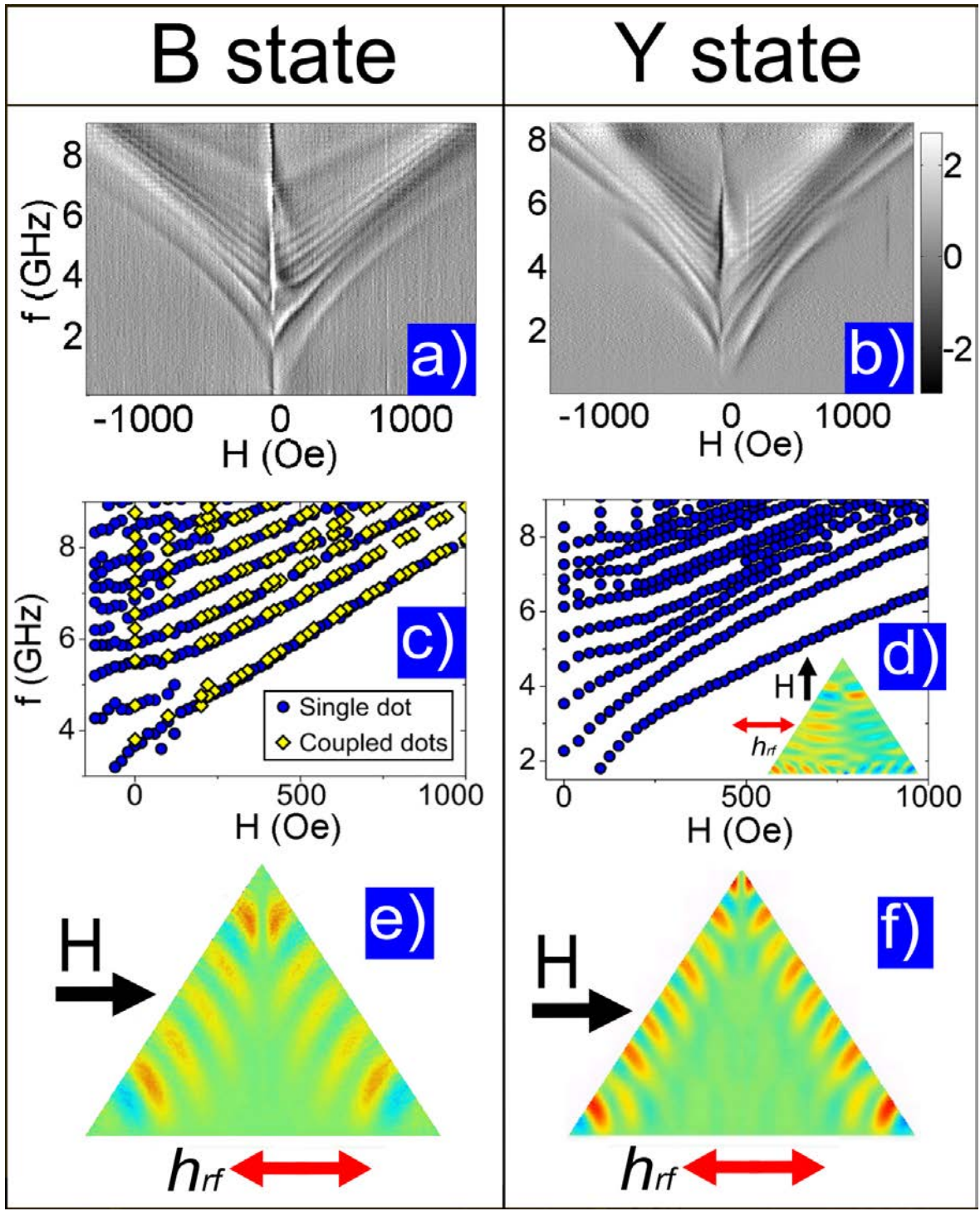

Figure 2. VNA-FMR measurements in the B state a), and in the Y state b) as a function of $H_{DC}$ (swept from positive to negative values, as indicated by gray arrows) and microwave frequency. Simulated eigenfrequencies of a single dot (blue points) and an array of 4x4 coupled dots (yellow points) for the B state c) and for a single dot only, in the Y state d). Inset in d) shows a snapshot of $M_X$ oscillations at f=12.46 GHz and $H_{DC}$=1000 Oe in the Y state. Snapshot of oscillations of $M_X$ in the B state at $H_{DC}$=500 Oe, at f=8.7 GHz e) and $H_{DC}$=1000 Oe, at f=12.46 GHz f).

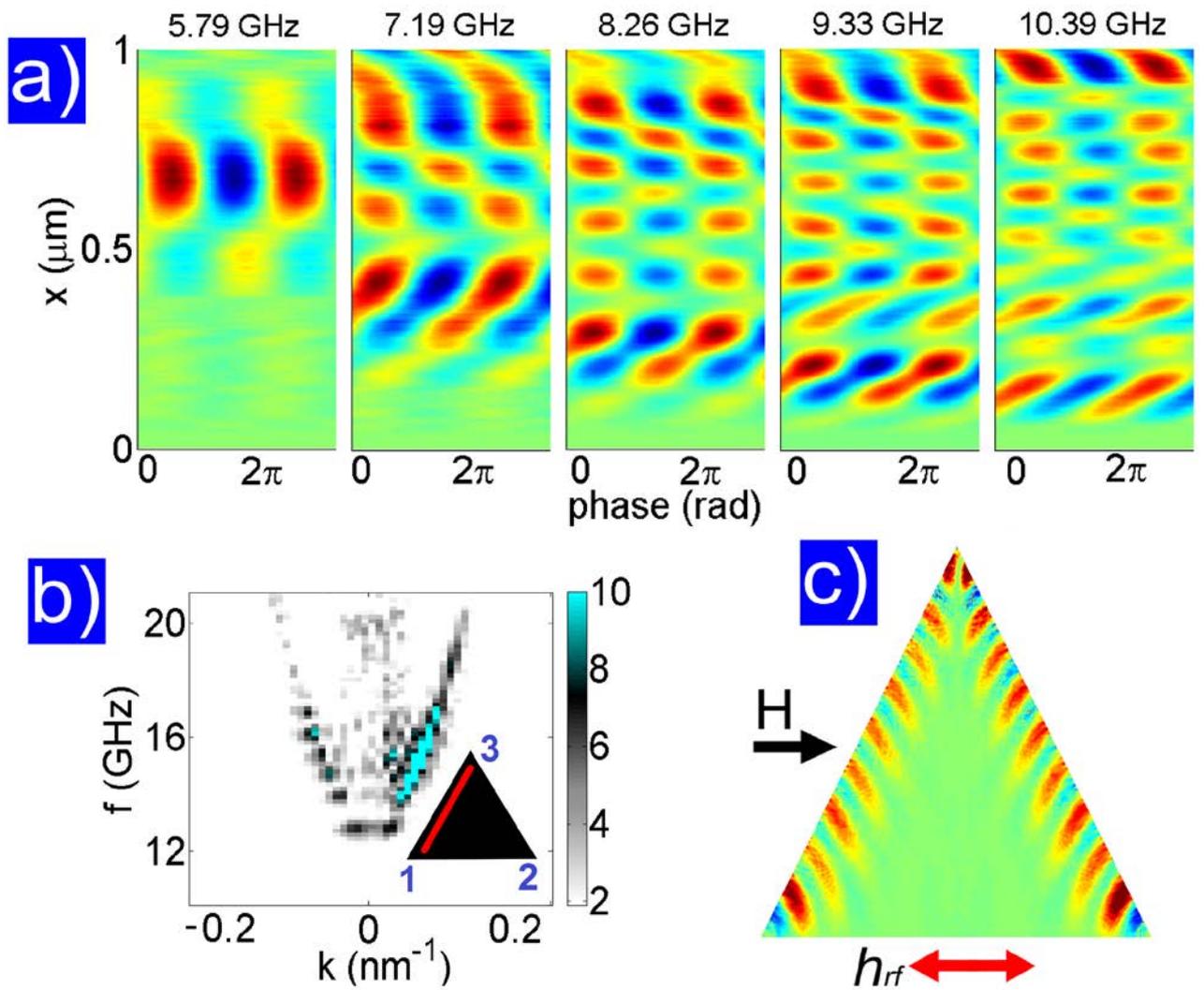

Figure 3. a) Simulated time evolution of edge spin waves along the line indicated in inset of b), for modes of different frequencies, with increasing number of nodes, during 1.5 oscillation periods. b) Simulated parabolic dispersion relation of edge spin waves ($M_Z$ component) along the line indicated in inset sketch from vertex 1 to 3. c) Edge spin waves ($M_X$) in an isosceles triangle (base = height = 1000 nm), similar to the equilateral case, for f=11.79 GHz, at $H_{DC}$=1000 Oe.

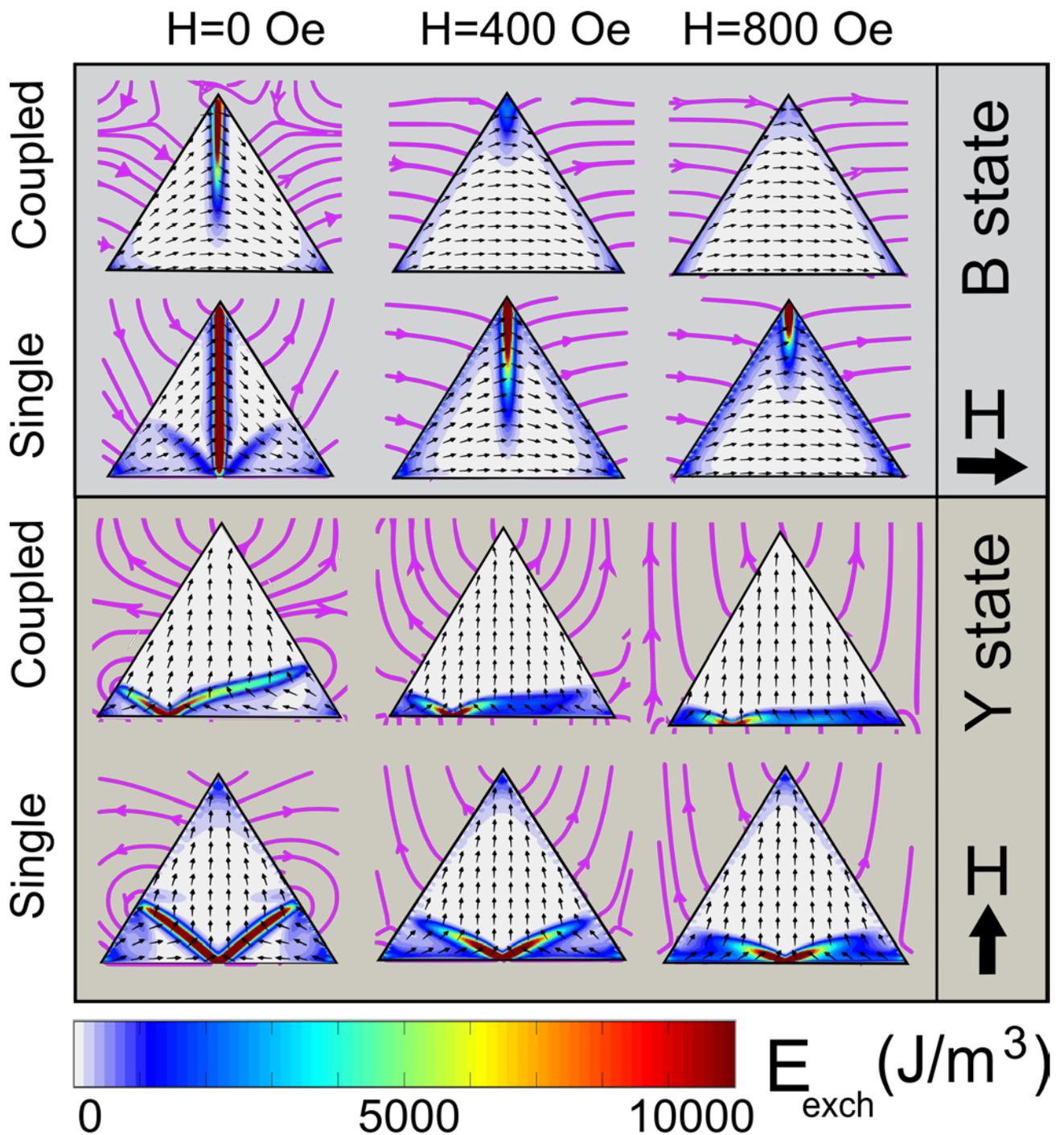

Figure 4. Static simulations comparing the field distribution (purple lines with arrows) and magnetic state (small black arrows for magnetization, colour scale for exchange energy density) in the case of an individual dot, or a dot laterally coupled to other dots, for different applied fields (indicated at top, in Oe). The top two rows represent the B state, and the bottom two rows represent the Y state, with and without considering coupling with neighbouring dots.

15